\documentclass[twocolumn,twocolappendix,tighten,dvipsnames]{aastex631}

\usepackage{amsmath,amssymb,amsfonts,slashed,bm,graphicx,braket}

\newcommand{\ev}[1]{{\left \langle {#1} \right \rangle}}
\newcommand{\var}{{\mathrm{Var}}}
\newcommand{\cov}{{\mathrm{Cov}}}
\newcommand{\corr}{{\mathrm{Corr}}}
\newcommand{\abs}[1]{{\left \lvert {#1} \right \rvert}}

\newcommand{\st}{\sigma_T}
\newcommand{\Dt}{{\Delta t}}
\newcommand{\DT}{{\Delta T}}
\newcommand{\mt}{\mu_T}

\shorttitle{Windowed Gaussian Processes}
\shortauthors{Lee and Gammie}

\begin{document}
	
\title{Measures of Variance on Windowed Gaussian Processes}

\author[0000-0002-3350-5588]{Daeyoung Lee}
\affiliation{Department of Physics, University of Illinois, 1110 West Green Street, Urbana, IL 61801, USA}

\author[0000-0001-7451-8935]{Charles F. Gammie}
\affiliation{Department of Physics, University of Illinois, 1110 West Green Street, Urbana, IL 61801, USA}
\affiliation{Department of Astronomy, University of Illinois at Urbana-Champaign, 1002 West Green Street, Urbana, IL 61801, USA}
\affiliation{NCSA, University of Illinois, 1205 W. Clark St., Urbana, IL 61801, USA}
\affiliation{Illinois Center for the Advanced Study of the Universe, University of Illinois, 1110 West Green St., Urbana, IL 61801, USA}

\begin{abstract}
	
	The variance and fractional variance on a fixed time window (variously known as ``rms percent'' or ``modulation index'') are  commonly used to characterize the variability of astronomical sources.  We summarize properties of this statistic for a Gaussian process.  
	
\end{abstract}

\section{Introduction}

A recent study of millimeter variability of Sgr A* \citep{wie22} makes use of the so-called modulation index $M$ on a fixed time window.  This statistic, which simply calculates the ratio between the standard deviation and the mean of the light curve on the window, has been used in various contexts (cf. \citet{vau03}, where it is called $F_\mathrm{var}$, the ``fractional rms variability amplitude''). The Event Horizon Telescope used this statistic to compare observed light curves with simulations \citep{eht}, but in the course of that work we were unable to find a clear summary of the effects of the length of the window on $M$ for Gaussian processes.  The relevant quantities are easy to derive; in this note we collect them in one place using consistent notation. 

In this brief note $\mu_T$ and $\sigma_T^2$ are the mean and variance (for example of a light curve) measured in a window of finite duration $T$.  Assuming the underlying time series is well modeled as a Gaussian process, we answer the following questions: (1) how is the covariance related to the expected $\sigma_T^2$? (2) given the covariance, what is the expected variance in $\sigma_T^2$ for a single measurement?  (3) how correlated are successive samples of $\sigma_T^2$ and, in particular, can they be treated as independent measurements? (4) what is the relationship of the expected $\sigma_T^2$ to the structure function? (5) what are the answers to these general questions for the important special case of a damped random walk? (6) what are the effects of sparse and nonuniform sampling?  (7) what are the implications for the modulation index?  In this discussion we neglect measurement errors. We also assume the observation duration is small compared to the window duration. 

\section{Measured Variance}

Let $f(t)$ be a stationary Gaussian random process with mean $\ev{f(t)} = \mu$ and unconditional variance $\ev{f^2(t)} - \mu^2 = \sigma^2$. Since the process is stationary, we may set the covariance function to be $\ev{f(t_1)f(t_2)} - \ev{f(t_1)}\ev{f(t_2)} \equiv \cov(t_1,t_2) = \cov(\abs{t_1 - t_2})$. Here $\ev{ \,\cdot\, }$ denotes the expectation value. 
    
When observed on a window $[0,T]$, the measured mean and variance are
\begin{align}
    \mt &\equiv \frac{1}{T} \int_0^T f(t) \,dt, \\
    \st^2 &\equiv \frac{1}{T} \int_0^T \left( f(t) - \mt \right)^2 \,dt.
\end{align}
Notice that while $\ev{\mt} = \mu$, $\mt \neq \mu$ but is rather a random variable that depends on $f(t)$. Let 
\begin{equation}
    \rho(t_1,t_2) \equiv \frac{1}{\sigma^2} \cov(t_1,t_2)
\end{equation}
be the correlation function of $f$, and define
\begin{equation}
    c(T) \equiv \frac{1}{T^2} \int_0^T \int_0^T \rho(t_1,t_2) \,dt_2 \,dt_1. \label{eq:ct}
\end{equation}
Then
\begin{align}
    \ev{\mt^2} &= \mu^2 + \sigma^2 c(T), \label{eq:evmtsq}\\
    \ev{\st^2} &= \sigma^2 (1 - c(T)). \label{eq:evstsq}
\end{align}

What is the variance of $\sigma_T^2$ for a single measurement?
It is helpful to define 
\begin{align}
    c_1(T) &\equiv \frac{1}{T^2} \iint \rho(t_1,t_2)^2 \,dt_2 \,dt_1, \\
    c_2(T) &\equiv \frac{1}{T^3} \iiint \rho(t_1,t_2) \rho(t_2,t_3) \,dt_3 \,dt_2 \,dt_1, \\
    c_3(T) &\equiv \frac{1}{T^4} \left[ \iint \rho(t_1,t_2) \,dt_2 \,dt_1 \right]^2 \nonumber\\
    &= [c(T)]^2, 
\end{align}
where all integrals are evaluated from $0$ to $T$. Then the variance of $\st^2$ is
\begin{equation}
    \var(\st^2) = 2 \sigma^4 \left[ c_1(T) - 2c_2(T) + c_3(T) \right].
\end{equation}

As a result of the correlation between points in a time series, the behavior of the measured mean and variance are dependent on both the covariance function and the length of the observation window. Assuming $c(T) \rightarrow 0$ as $T \rightarrow \infty$, $\ev{\st^2}$ is smaller than the unconditional variance $\sigma^2$ for small $T$, and approaches it as $T$ increases. 

\section{Correlation Between Different Windows}

How independent are measurements of $\sigma_T^2$ over successive windows?  Consider a second window $[\DT, T + \DT]$, with
\begin{align}
    \mt' &\equiv \frac{1}{T} \int_\DT^{T+\DT} f(t) \,dt, \\
    \st'^2 &\equiv \frac{1}{T} \int_\DT^{T+\DT} \left( f(t) - \mt \right)^2 \,dt.
\end{align}
Since $f$ is stationary and thus $\rho(t_1,t_2) = \rho(\abs{t_2-t_1})$, define
\begin{align}
    c'(T) &\equiv \frac{1}{T^2} \int_0^T \int_\DT^{T+\DT} \rho(t_1,t_2) \,dt_2 \,dt_1, \\
    c'_1(T) &\equiv \frac{1}{T^2} \int_0^T \int_\DT^{T+\DT}  \rho(t_1,t_2)^2 \,dt_2 \,dt_1, \\
    c'_2(T) &\equiv \frac{1}{T^3} \int_0^T \int_0^T \int_\DT^{T+\DT} \rho(t_1,t_3) \rho(t_2,t_3) \,dt_3 \,dt_2 \,dt_1 \nonumber\\
    = \frac{1}{T^3} & \int_\DT^{T+\DT} \int_\DT^{T+\DT} \int_0^T \rho(t_1,t_3) \rho(t_2,t_3) \,dt_3 \,dt_2 \,dt_1, \\
    c'_3(T) &\equiv \frac{1}{T^4} \left[ \int_0^T \int_\DT^{T+\DT} \rho(t_1,t_2) \,dt_2 \,dt_1 \right]^2 \nonumber\\
    &= [c'(T)]^2.
\end{align}
Then
\begin{equation}
    \cov(\st^2, \st'^2) = 2 \sigma^4 \left[ c'_1(T) - 2c'_2(T) + c'_3(T) \right].
\end{equation}
Evidently the measured variance on windows with a small time separation are correlated. The degree of correlation depends on the separation between windows, the length of the windows, and the shape of the covariance function. 

\section{Relationship Between \texorpdfstring{$\st^2$} and the Structure Function}

The structure function at a time lag $\Dt$, here defined as 
\begin{equation}
    SF(\Dt) = \frac{1}{T-\Dt} \int_0^{T-\Dt} \left[ f(t+\Dt) - f(t) \right]^2 \,dt
\end{equation}
is closely related to $\sigma_T^2$, since
\begin{equation}
    \ev{ SF(\Dt) } = 2\sigma^2 - 2 \mathrm{Cov}(\Dt)
\end{equation}
and thus
\begin{equation}
    \ev{ \st^2 } = \frac{1}{T^2} \int_0^T \int_0^T \frac{\ev{SF(\abs{t_1 - t_2})}}{2} \,dt_1 \, dt_2.
\end{equation}
Since $c(y)$ corresponds to the integral of the covariance function over $[0,T] \times [0,T]$, $\ev{\sigma_T^2}$ corresponds to the integral of the structure function over $[0,T] \times [0,T]$. 

\section{Damped Random Walk Example}

For the damped random walk
\begin{align}
    \cov(t_1,t_2) &= \sigma^2 e^{ -\abs{t_1 - t_2} / \tau }, \\
    SF(\Dt) &= 2 \sigma^2 \left( 1 - e^{-\Dt / \tau} \right).
\end{align}
Recall that $\Delta t > 0$ is a time lag and is not the same as $\Delta T$, which labels differences between windows. For $y \equiv T / \tau$,
\begin{align}
    c(y) &= 2 \left( \frac{1}{y} - \frac{1}{y^2} (1 - e^{-y}) \right) ,\\
    c_1(y) &= c(2y), \\
    c_2(y) &= \frac{1}{y^3} \left( 2y (2 + e^{-y}) - (7 - e^{-y}) (1 - e^{-y}) \right), \\
    c_3(y) &= [c(y)]^2,
\end{align}
with 
\begin{align}
    \ev{\st^2} &= \sigma^2 (1-c(y)), \\
    \var(\st^2) &= 2 \sigma^4 \left[ c(2y) - 2c_2(y) + c^2(y) \right].
\end{align}

For illustrative purposes we will look at the correlation between consecutive windows, i.e. taking $\Delta T = T$. Then, 
\begin{align}
    c'(y) &= \frac{1}{y^2} (1 - e^{-y})^2, \\
    c_1'(y) &= c'(2y), \\
    c_2'(y) &= \frac{1}{2y^3} (1 - e^{-y})^3 (1 + e^{-y}), \\
    c_3'(y) &= [c(y)]^2,
\end{align}
with
\begin{equation}
    \cov(\st^2, \st'^2) = 2 \sigma^4 \left[ c'(2y) - 2c'_2(T) + c'^2(y) \right],
\end{equation}
and the correlation between the two segments is
\begin{equation}
    \corr(\st^2, \st'^2) = \frac{\cov(\st^2, \st'^2)}{\var(\st^2)}.
\end{equation}

As $y \rightarrow 0$, 
\begin{align}
    \ev{\st^2} &\rightarrow \sigma^2 \left( \frac{y}{3} - \frac{y^2}{12} + \mathcal{O}(y^3) \right), \\
    \var(\st^2) &\rightarrow \sigma^4 \left( \frac{4y^2}{45} - \frac{y^3}{15} + \mathcal{O}(y^4) \right), \\
    \cov(\st^2, \st'^2) &\rightarrow \sigma^4 \left( \frac{y^4}{72} - \frac{y^5}{36} + \mathcal{O}(y^6) \right), \\
    \corr(\st^2, \st'^2) &\rightarrow \frac{5y^2}{32} - \frac{25y^3}{128} + \mathcal{O}(y^4),
\end{align}
and as $\frac{1}{y} \rightarrow 0$, 
\begin{align}
    \ev{\st^2} &\rightarrow \sigma^2 \left( 1 - \frac{2}{y} + \mathcal{O}(y^{-2}) \right), \\
    \var(\st^2) &\rightarrow \sigma^4 \left( \frac{2}{y} - \frac{9}{y^2} + \mathcal{O}(y^{-3}) \right), \\
    \cov(\st^2, \st'^2) &\rightarrow \sigma^4 \left( \frac{1}{2y^2} - \frac{2}{y^3} + \mathcal{O}(y^{-4}) \right), \\
    \corr(\st^2, \st'^2) &\rightarrow \frac{1}{4y} + \frac{1}{8y^2} + \mathcal{O}(y^{-3}).
\end{align}
All four quantities are plotted in Figure \ref{fig:plot}. Evidently  the correlation between consecutive windows disappears as $T \rightarrow 0$ or $T \rightarrow \infty$, and peaks at $T \approx 2 \tau$. 

\section{Sparse and Non-Uniform Sampling}

In the previous section, $\st^2$ was evaluated as an integral. For real data, if the observation was uniformly and densely sampled (relative to both the characteristic timescale $\tau$ and the length of the observation $T$), $\st^2$ can be approximated as the variance of the observation. However, data is commonly sampled sparsely and nonuniformly. 

If the observation is sampled by $f_i = f(t_i)$, $i = 1, \cdots, N$, then we would measure 
\begin{align}
    \mt &= \frac{1}{N} \sum_{i=1}^N f_i, \\
    \st^2 &= \frac{1}{N} \sum_{i=1}^N \left[ f_i - \mt \right]^2.
\end{align}
Then
\begin{align}
    \ev{\mt} &= \mu, \\
    \ev{\mt^2} &= \mu^2 + \frac{\sigma^2}{N^2} \sum_{i=1}^N \sum_{j=1}^N \rho(t_i, t_j), \\
    \ev{\st^2} &= \sigma^2 \left[ 1 - \frac{1}{N^2} \sum_{i=1}^N \sum_{j=1}^N \rho(t_i, t_j) \right], 
\end{align}
are the discrete analogues of equations \ref{eq:evmtsq} and \ref{eq:evstsq}. If the sampling was uniform, the sum would serve as an approximation to the integral in equation \ref{eq:ct}, but non-uniform sampling will bias our estimate. The nature of this bias depends on the details of the sampling. 

One strategy for minimizing bias would be to weight each sample by the gaps $\Dt_i$ between observations. However, this can fail if there are large gaps in the data. For example, consider a densely sampled observation across the first and third hour of a three-hour observation, with an hour-long gap in the middle of the observation period. On the first and third segments of observation, we can assume that the measured variance of the segments well-approximates the integrated variance of the underlying process. 

For this example, we can look at the effects of two different methods of addressing this gap in our observation: taking the unweighted variance across our samples and weighting each sample by the time gaps in observation before and after the sample. 

For a given statistic $g(t)$ calculated from  $f(t)$ (such as the point-wise variance, $g(t) = f(t)^2 - \mt^2$), the underlying ``true'' measured value from $t=0h$ to $t=3h$ is
\begin{equation}
    G = \frac{1}{3} \int_0^3 g(t) \,dt.
\end{equation}
The first method essentially assumes all observations are uniformly spaced across the entire observation, dilating the function $g(t)$:
\begin{align}
    G_1 &= \frac{1}{3} \left[ \int_0^{3/2} g(2t/3) \,dt + \int_{3/2}^3 g(2t/3+1) \,dt \right] \\
    &= \frac{1}{2} \left[ \int_0^1 g(t) \,dt + \int_2^3 g(t) \,dt \right].
\end{align}
For a quantity like the variance, this gives a consistent underestimate of the true measured value across three hours. On the other hand, the second method instead places undue weight on the endpoints:
\begin{equation}
    G_2 = \frac{1}{3} \left[ \int_0^1 g(t) \,dt + \frac{1}{2} g(1) + \frac{1}{2} g(2) + \int_1^2 g(t) \,dt \right].
\end{equation}
The resulting value of $G_2$ is strongly dependent on the values $g(1)$ and $g(2)$ at the endpoints of the gap in the observation. 

Thus, it is useful to Monte Carlo the analysis on Gaussian processes with the same sampling as the data, to see the effects of the sampling on the measurement of variability. In our applications the variance in the measurement tends to outweigh any effects of sampling. 

\section{Modulation Index}

The modulation index or rms percent is $M_T \equiv \st / \mt$, the standard deviation divided by the mean over a period of observation. For a Gaussian process, the measured variance and the measured mean are uncorrelated, since
\begin{align*}
    \ev{\st^2 \mt} &= \ev{ \frac{1}{T} \int_0^T \left[ f^2(t) \mt - \mt^3 \right] \,dt } \\
    &= \frac{1}{T^2} \iint \ev{f_1^2 f_2} \,dt_1 \,dt_2 \nonumber\\
    &\qquad - \frac{1}{T^3} \iint \ev{f_1 f_2 f_3} \,dt_1 \,dt_2 \,dt_3 \\
    &= 0,
\end{align*}
The standard deviation and mean are similarly uncorrelated. However, they are clearly not independent. Thus, rather than deriving an analytic expression for $M_T$, we will explore it computationally. 

\begin{figure*}[h]
    \centering
    \includegraphics[width=0.8\textwidth]{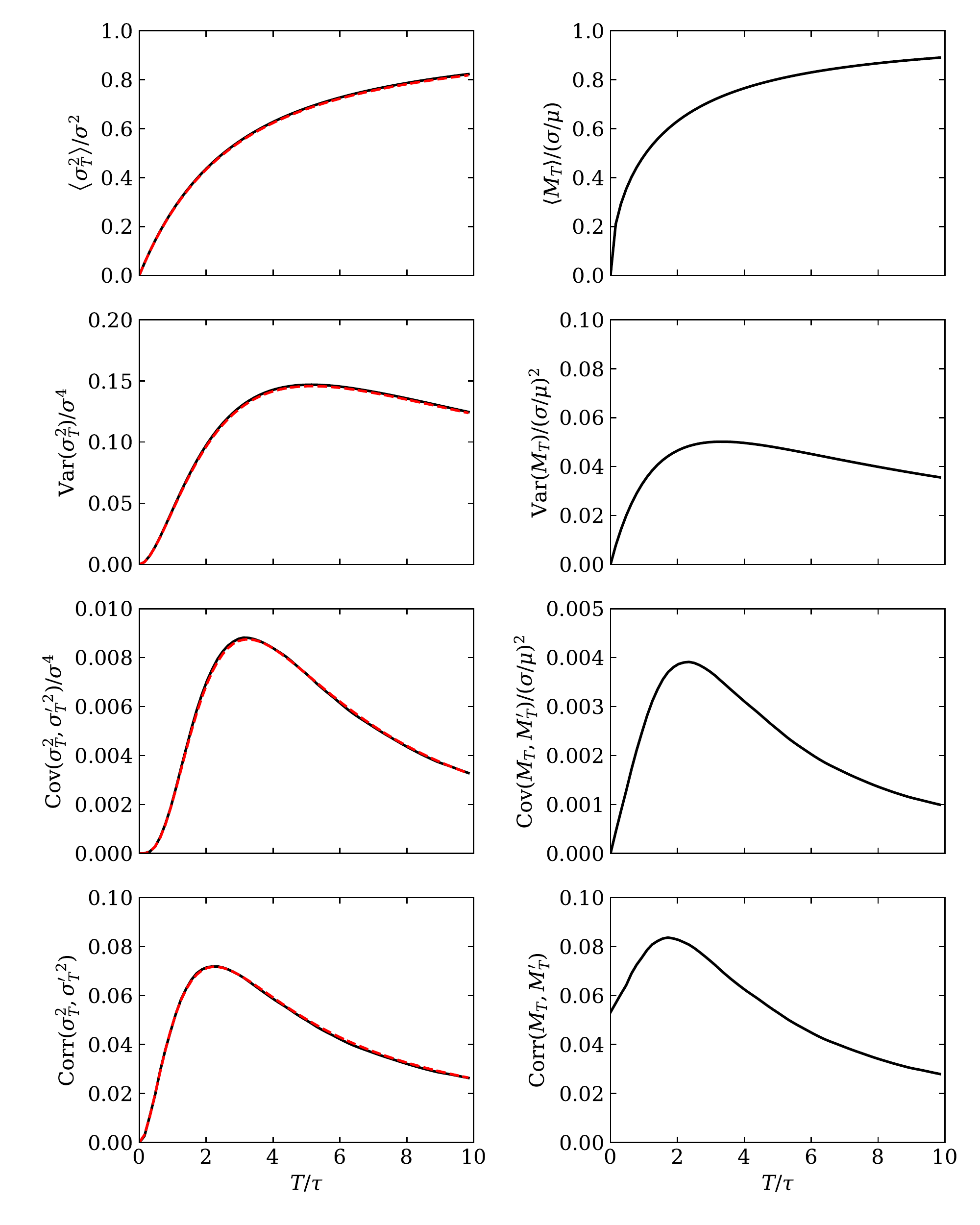}
    \caption{The mean and variance of $\st^2$ and $M_T$ for a damped random walk, along with the covariance and correlation of these measurements on consecutive windows, as functions of the window length $T / \tau$, averaged over $10^7$ realizations with $\sigma = 0.24$ and $\mu = 2.4$. Analytic results, where available, are shown in red. The means are normalized by $\sigma^2$ and $\sigma / \mu$, respectively, and the variance and covariance are normalized by $\sigma^4$ and $(\sigma / \mu)^2$.  \label{fig:plot}}
\end{figure*}

Figure \ref{fig:plot} shows the mean and variance of $M_T$ for a damped random walk, along with the covariance and correlation of measurements of $M_T$ between consecutive windows. Similar statistics for $\st^2$ are also given. These quantities were averaged over $10^7$ independent realizations, with damped random walk parameters $\sigma = 0.24$, $\mu = 2.4$, and $\tau = 1$. Each damped random walk was generated on 2048 uniformly spaced points between $0$ and $20$, and the statistics were measured as a function of the window length $T \in [0,10]$, calculated on the first window (or first two, where applicable).
    
Qualitatively, $\st^2$ and $M_T$ behave similarly. Of particular interest is the correlation between consecutive windows, which peaks on the order of $\tau$ but remains fairly low across all window lengths, never rising above $0.1$. Thus, as long as $\st^2$ and $M_T$ are measured on independent windows ($\DT \geq T$), the correlation between measurements remains small. 

\begin{acknowledgements}
This work was supported by NSF grants AST 17-16327 (horizon), OISE 17-43747, and AST 20-34306, and by the Grainger College of Engineering's Willett Chair.
\end{acknowledgements}

\end{document}